\documentclass[letterpaper]{article} % DO NOT CHANGE THIS
\usepackage{aaai24}  % DO NOT CHANGE THIS
\usepackage{times}  % DO NOT CHANGE THIS
\usepackage{helvet}  % DO NOT CHANGE THIS
\usepackage{courier}  % DO NOT CHANGE THIS
\usepackage[hyphens]{url}  % DO NOT CHANGE THIS
\usepackage{graphicx} % DO NOT CHANGE THIS
\usepackage{natbib}  % DO NOT CHANGE THIS AND DO NOT ADD ANY OPTIONS TO IT
\usepackage{caption} % DO NOT CHANGE THIS AND DO NOT ADD ANY OPTIONS TO IT
\usepackage{algorithm}
\usepackage{algorithmic}
\usepackage{amsmath}
\usepackage{amsthm}
\usepackage{amssymb}
\usepackage{amsfonts}

\usepackage{tabularx}
\usepackage{booktabs}   % Recommended in nips/icml, ``for professional tables'', enables toprule/midrule/bottomrule.
\usepackage{multirow}
\usepackage{subfigure}

\usepackage{manfnt}             % for \dbend

\usepackage{xspace}             % \xspace becomes a space in word boundary
\usepackage{relsize}            % scaling text size
\usepackage{adjustbox}          % adjusting a box width/height
\usepackage{diagbox}            % for diagonal crossing in tables
\usepackage{pdflscape}          % for landscape pages in appendix

\usepackage{microtype}

\usepackage{graphicx}
\usepackage{url}
\usepackage{comment}            % for comment environment
\usepackage{xcolor}             % see patch.sty for option quirks
\usepackage{colortbl}
\newcommand*\samethanks[1][\value{footnote}]{\footnotemark[#1]}

\usepackage{newfloat}
\usepackage{listings}
\DeclareCaptionStyle{ruled}{labelfont=normalfont,labelsep=colon,strut=off} % DO NOT CHANGE THIS
\floatstyle{ruled}
\newfloat{listing}{tb}{lst}{}
\floatname{listing}{Listing}
\title{MedSegDiff-V2: Diffusion based Medical Image Segmentation with Transformer}

\author{
Junde Wu\textsuperscript{\rm 1,6,7,8},
Wei Ji\textsuperscript{\rm 2},
Huazhu Fu\textsuperscript{\rm 3},
Min Xu\thanks{co-corresponding author}, \textsuperscript{\rm 4,6}\,
Yueming Jin\textsuperscript{\rm 1},
Yanwu Xu\samethanks \textsuperscript{\rm 5},
}
\affiliations {
\textsuperscript{\rm 1}National University of Singapore,
\textsuperscript{\rm 2}University of Alberta,
\textsuperscript{\rm 3}A*STAR,
\textsuperscript{\rm 4}Carnegie Mellon University,
\textsuperscript{\rm 5}Singapore Eye Research Institute,
\textsuperscript{\rm 6}Mohamed bin Zayed University of Artificial Intelligence,
\textsuperscript{\rm 7}University of Oxford,
\textsuperscript{\rm 8}Kids with Tokens\\
jundewu@ieee.org, 
ywxu@ieee.org,
xumin100@gmail.com
}
\begin{document}

\maketitle

\begin{abstract}
The Diffusion Probabilistic Model (DPM) has recently gained popularity in the field of computer vision, thanks to its image generation applications, such as Imagen, Latent Diffusion Models, and Stable Diffusion, which have demonstrated impressive capabilities and sparked much discussion within the community. Recent investigations have further unveiled the utility of DPM in the domain of medical image analysis, as underscored by the commendable performance exhibited by the medical image segmentation model across various tasks. Although these models were originally underpinned by a UNet architecture, there exists a potential avenue for enhancing their performance through the integration of vision transformer mechanisms. However, we discovered that simply combining these two models resulted in subpar performance. To effectively integrate these two cutting-edge techniques for the Medical image segmentation, we propose a novel Transformer-based Diffusion framework, called MedSegDiff-V2. We verify its effectiveness on 20 medical image segmentation tasks with different image modalities. Through comprehensive evaluation, our approach demonstrates superiority over prior state-of-the-art (SOTA) methodologies. Code is released at \url{https://github.com/KidsWithTokens/MedSegDiff}

\end{abstract}
%
% discussed a lot about model design with ChatGPT 
% writing rephrased a lot by ChatGPT
 
\section{Introduction}
 
Medical image segmentation is to divide a medical image into distinct regions of interest. It is a crucial step in many medical applications, such as diagnosis and image-guided surgery. In recent years, there has been a growing interest in automated segmentation methods, as they have the potential to improve the consistency and accuracy of results. With the advancement of deep learning techniques, several studies have successfully applied neural network-based models, including classical convolutional neural networks (CNNs) \cite{ji2021learning, wu2022seatrans} and the recently popular vision transformers (ViTs)\cite{chen2021transunet, wang2021transbts}, to medical image segmentation tasks.

Very recently, the Diffusion Probabilistic Model (DPM)\cite{ho2020denoising} has gained popularity as a powerful class of generative models, capable of generating high-quality and diverse images\cite{ramesh2022hierarchical, saharia2022photorealistic, rombach2022high}. Inspired by its success, many researches have applied DPM in the field of medical image segmentation\cite{wu2022medsegdiff, wolleb2021diffusion, kim2022diffusion, guo2022accelerating, rahman2023ambiguous}. Many of them reported new SOTA on several benchmarks by using the DPM. The remarkable performance of this model stems from its inherent stochastic sampling process\cite{wu2022medsegdiff, rahman2023ambiguous}.  DPM has the capability to generate different segmentation predictions by running multiple times. The diversity among these samples directly captures the uncertainty associated with targets in medical images, where organs or lesions commonly have ambiguous boundaries. However, it is worth noting that all these methods rely on classical UNet backbones. In comparison to the increasingly popular vision transformers, classical UNet models compromise on segmentation quality, which can lead to the generation of divergent yet incorrect masks in ensemble, ultimately introducing noise that permanently hampers the performance.

A natural next step is to combine the transformer-based UNet, such as TransUNet\cite{chen2021transunet}, with DPM. However, we found that implementing it in a straightforward manner resulted in subpar performance. One issue is that the transformer-abstracted conditional feature is not compatible with the feature of the diffusion backbone. The transformer is able to learn deep semantic features from the raw image, whereas the diffusion backbone abstracts features from a corrupted and noisy mask, making feature fusion more challenging. Additionally, the dynamic and global nature of the transformer makes it more sensitive than CNNs \cite{naseer2021intriguing}. Thus, the adaptive condition strategy used in previous diffusion-based methods\cite{wu2022medsegdiff} will cause large variance in the transformer setting. This leads more ensemble and converge difficulties.

\begin{figure*}
    \centering
    \includegraphics[width=0.8\linewidth]{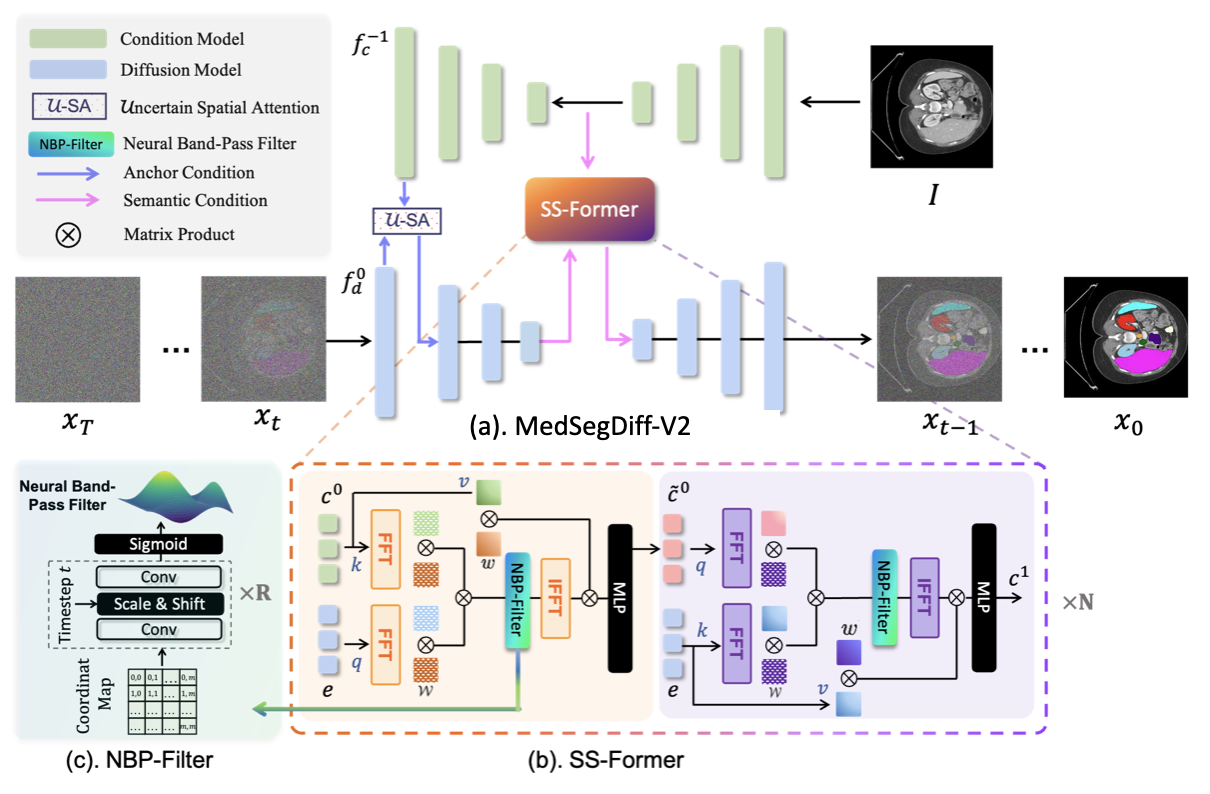}
    \caption{An illustration of MedSegDiff-V2, which starts from (a) an overview of the pipeline, and continues with zoomed-in diagrams of individual Models, including (b) SS-Former, and (c) NBP-Filter.}
    \label{fig:framework}
\end{figure*}

To overcome the aforementioned challenges, we have designed a novel Transformer-based Diffusion framework for the Medical image segmentation, called MedSegDiff-V2. The main idea is to employ two different conditioning techniques over the backbone with the raw image in the diffusion process. One is the Anchor Condition, which integrates the conditional segmentation features into the diffusion model encoder to reduce the diffusion variance. We design a novel $\mathcal{U}$ncertain Spatial Attention ($\mathcal{U}$-SA) mechanism for the integration, which relaxes the conditional segmentation feature with more uncertainty, thus providing the diffusion process more flexibility to further calibrate the predictions. The other is the Semantic Condition that integrates the conditional embedding into the diffusion embedding. To effectively bridge the gap between these two embedding, we propose a novel transformer mechanism called the Spectrum-Space Transformer (SS-Former) for the embedding integration. SS-Former is a cross-attention chain in frequency domain, with a timestep-adaptive Neural Band-pass Filter (NBP-Filter) to align the noise and semantic features each time. 

In brief, the contributions of this paper are: 
\begin{itemize}
    \item We are the first to integrate transformer into a diffusion-based model for general medical image segmentation. 
    
    \item We propose an Anchor Condition with $\mathcal{U}$-SA to mitigate the diffusion variance. 
    
    \item We propose Semantic Condition with SS-Former to model the segmentation noise and semantic feature interaction. 
    
    \item We achieve SOTA performance on 20 organ segmentation tasks including 5 image modalities.
\end{itemize}
% 1). The first to integrate transformer into a diffusion-based model for general medical image segmentation. 2). An Anchor Condition with $\mathcal{U}$-SA to mitigate the diffusion variance. 3). A Semantic Condition with SS-Former to model the segmentation noise and semantic feature interaction. 4). SOTA performance on five medical segmentation benchmarks.
%Contributions
% \begin{itemize}
% \item The first to integrate transformer into a diffusion-based model for general medical image segmentation.

% \item An Anchor Condition with $\mathcal{U}$-SA to mitigate the diffusion variance.

% \item A Semantic Condition with SS-Former to model the segmentation noise and semantic feature interaction.

% \item SOTA performance on  on 18 organ segmentation tasks including different image modalities. 
% \end{itemize}
\section{Related Work}
\subsection{Medical Segmentation with Transformers}
% Vision transformers
% % , initially derived from sequence-to-sequence models in natural language processing, utilize self-attention mechanisms to aggregate information from input sequences. They 
% have demonstrated comparable or superior performance compared to convolutional architectures such as ResNet\cite{resnet} and U-Net\cite{3dunet}. 
Previous studies have highlighted the potential of transformer-based models to achieve SOTA results in medical image segmentation. One notable example is TransUNet\cite{chen2021transunet}, which combined the transformer with UNet as a bottleneck feature encoder. Since then, several works have proposed incorporating cutting-edge transformer techniques into the backbone of medical image segmentation models, including Swin-UNet\cite{cao2022swin}, Swin-UNetr\cite{swin-unetr}, and DS-TransUNet\cite{lin2022ds}. As recently UNet based diffusion-based segmentation models have recently emerged as achieving new SOTA in medical image segmentation, it is worthwhile to explore ways to integrate recognized transformer architectures into this powerful new backbone.

\subsection{Diffusion Model for Medical Segmentation}
Diffusion models have recently demonstrated significant potential in various segmentation tasks, including medical images \cite{armato2011lung, caron2021emerging, cao2022swin, chen2019med3d}. In fact, these models leverage a stochastic sampling process to generate an implicit ensemble of segmentations, leading to enhanced segmentation performance \cite{zhai2022scaling}. However, without effective control of diversity, the ensemble often struggles to converge
% , resulting in multiple time-consuming sampling iterations. Moreover, these divergent samples not only fail to meet the desired target but also introduce noise that hampers segmentation quality. 
Therefore, it is crucial to improve the sample accuracy with each sampling iteration.

\section{Method}
\subsection{Diffusion process of MedSegDiff-V2}
We have designed our model based on the diffusion model mentioned in \cite{ho2020denoising}. Diffusion models are generative models that consist of two stages: a forward diffusion stage and a reverse diffusion stage. In the forward process, Gaussian noise is gradually added to the segmentation label $x_{0}$ through a series of steps $T$. In the reverse process, a neural network is trained to recover the original data by reversing the noise addition process. This can be mathematically represented as follows:
\begin{equation}
p_{\theta}(x_{0:T-1}|x_{T}) = \prod_{t=1}^{T} p_{\theta}(x_{t-1}|x_{t}),
\end{equation}
where $\theta$ represents the parameters of the reverse process. Starting from a Gaussian noise distribution, $p_{\theta}(x_{T}) = \mathcal{N}(x_{T};0,I_{n \times n})$, where $I$ is the raw image, the reverse process transforms the latent variable distribution $p_{\theta}(x_{T})$ to the data distribution $p_{\theta}(x_{0})$. To maintain symmetry with the forward process, the reverse process recovers the noisy image step by step, ultimately obtaining the final clear segmentation.

Following the standard implementation of DPM, we utilize an encoder-decoder network for the learning. To achieve segmentation, we condition the step estimation function $\epsilon$ on the prior information from the raw image. This conditioning can be expressed as:
\begin{equation}
\epsilon_{\theta}(x_{t},I,t) = D(TransF(E_{t}^{I}, E_{t}^{x}),t),
\end{equation}
Here,  $TransF$ denotes the transformer based attention mechanism. $E_{t}^{I}$ represents the conditional feature embedding, which, in our case, corresponds to the embedding of the raw image. $E_{t}^{x}$ represents the feature embedding of the segmentation map for the current step. These two components are incorporated together by transformer and passed through a UNet decoder $D$ for reconstruction. The step index $t$ is integrated with the combined embedding and decoder features, and each step index is embedded using a shared learned look-up table, following the approach described in \cite{ho2020denoising}.

% The Diffusion Probabilistic Model (DPM) is a generative model composed of two stages, a forward diffusion stage and a reverse diffusion stage. In the forward process, the segmentation label $x_{0}$ is gradually added Gaussian noise through a series of steps $T$. In the reverse process, a neural network is trained to recover the original data by reversing the noising process. The reverse process starts with a Gaussian noise and transforms the latent variable distribution to the data distribution. A UNet is adopted as the network for learning, and the step estimation function is conditioned by raw image prior. In this way, the model can recover the noise image step by step to obtain the final clear segmentation.
  
\subsection{Overall architecture}\label{AA}
  
The overall flow of MedSegDiff-V2 is shown in \ref{fig:framework}. To introduce the process, consider a single step $t$ of the diffusion process. The noisy mask $x_{t}$ is first inputted to a UNet, called the Diffusion Model. Diffusion Model is conditioned by the segmentation features extracted from the raw images through another standard UNet, called the Condition Model. Two different conditioning manners are applied to the Diffusion Model: Anchor Condition and Semantic Condition. Following the flow of the input, the Anchor Condition is first imposed on the encoder of the Diffusion Model. It integrates the anchor segmentation features, which are the decoded segmentation features of the Condition Model, into the encoded features of the Diffusion Model. This allows the diffusion model to be initialized by a rough but static reference, which helps to reduce the diffusion variances. The Semantic Condition is then imposed on the embedding of the Diffusion Model, which integrates the semantic segmentation embedding of the Condition Model into the embedding of the Diffusion Model. This conditional integration is implemented by SS-Former, which bridges the gap between the noise and semantic embedding, and abstracts a stronger representation with the advantage of the global and dynamic nature of transformer.

MedSegDiff-V2 is trained using a standard noise prediction loss $\mathcal{L}_{n}$ following DPM\cite{ho2020denoising} and an anchor loss $\mathcal{L}_{anc}$ supervising the Condition Model. $\mathcal{L}_{anc}$ is a combination of soft dice loss $\mathcal{L}_{dice}$ and cross-entropy loss $\mathcal{L}_{ce}$. Specifically, the total loss function is represented as:
\begin{equation}
    \mathcal{L}_{total}^{t} = \mathcal{L}_{n}^{t} + (t \equiv 0\pmod{\alpha}) (\mathcal{L}_{dice} + \beta \mathcal{L}_{ce})
\end{equation}
where $t \equiv 0\pmod{\alpha}$ control the times of supervision over Condition Model through hyper-parameter $\alpha$, cross-entropy loss is weighted by hyper-parameter $\beta$ , which are set as 5 and 10 respectively.

%channel attention
  
\subsection{Anchor Condition with $\mathcal{U}$-SA}
  
Without the inductive bias of convolution layer, transformer blocks have stronger representation capability but are also more sensitive to the input variance when training data is limited\cite{naseer2021intriguing}. Directly adding the transformer block to the Diffusion Model will cause the large variance on each time outputs. To overcome this negative effect, we adapt the structure of MedSegDiff\cite{wu2022medsegdiff} and introduce the Anchor Condition operation to the Diffusion Model. 

Anchor Condition provides a rough anchor feature from the Condition Model and integrates it into the Diffusion Model. This provides the Diffusion Model with a correct range for predictions while also allowing it to further refine the results. Specifically, we integrate the decoded segmentation features of the Condition Model into the encoder features of the Diffusion Model. We propose $\mathcal{U}$-SA mechanism for the feature fusion to represent the uncertainty nature of the given conditional features. Formally, consider we integrate the last conditional feature $f_{c}^{-1}$ into the first diffusion feature $f_{d}^{0}$. $\mathcal{U}$-SA can be expressed as:
\vspace{-3pt}
\begin{align}
 & f_{anc} = Max(f_{c}^{-1} * k_{Gauss}, f_{c}^{-1}), \\
 & f_{d}^{'0} = Sigmoid (f_{anc} * k_{Conv_{1 \times 1} }) \cdot f_{d}^{0} + f_{d}^{0},
\end{align}\label{eqn:GSA1}

% \begin{equation}\label{eqn:GSA1}
%     f_{anc} = Max(f_{c}^{-1} * k_{Gauss}, f_{c}^{-1}), \\ and
% \end{equation}
% \begin{equation}\label{eqn:GSA2}
%     f_{d}^{'0} = Sigmoid (f_{anc} * k_{Conv_{1 \times 1} }) \cdot f_{d}^{0} + f_{d}^{0},
% \end{equation}
where $*$ denotes slide-window kernel manipulation, $\cdot$ denotes general element-wise manipulation. In the equation, we first apply a learnable Gaussian kernel $k_{G}$ over $f_{c}^{-1}$ to smooth the activation, as $f_{c}^{-1}$ serves as an anchor but may not be completely accurate. We then select the maximum value between the smoothed map and the original feature map to preserve the most relevant information, resulting in a smoothed anchor feature $f_{anc}$. Then we integrate $f_{anc}$ into $f_{d}^{0}$ to obtain an enhanced feature $f_{d}^{'0}$.
Specifically, we first apply a $1 \times 1$ convolution $k_{1 \times 1 conv}$ to reduce the anchor feature channels to $1$ and multiply it with $f_{d}^{0}$ after the Sigmoid activation, then add it to each channel of $f_{d}^{0}$, similar to the implementation of spatial attention\cite{woo2018cbam}.
% following the general implementation of spatial attention\cite{woo2018cbam}.
%Specifically, we first apply a $1 \times 1$ convolution $k_{1 \times 1 conv}$ to reduce the number of channels in the anchor feature to $1$. Then, we use a sigmoid activation function on the anchor feature and add it to each channel of $f_{d}^{0}$, similar to the implementation of spatial attention\cite{woo2018cbam}. 

\subsection{Semantic Condition with SS-Former}
  
% We propose a novel transformer architecture, called Spectrum-Space Transformer (SS-Former), to effectively integrate the conditional segmentation embedding into the diffusion embedding. SS-Former is composed of several blocks that share the same architecture. Each block consists of two cross-attention-like modules. The first encodes the diffusion noise embedding into the condition semantic embedding, and the next module encodes the noise-blended semantic embedding into the diffusion noise embedding. This allows the model to learn the interaction between noise and semantic features and achieve a stronger representation.
The Diffusion Model predicts redundant noise from a noisy mask input, leading to a domain gap between its embedding and the conditional segmentation semantic embedding. This divergence compromises performance when using matrix manipulations, such as in a stranded transformer. To address this challenge, we propose a novel Spectrum-Space Transformer (SS-Former). Our key idea is to learn the interaction of condition semantic feature and diffusion noise feature in the frequency domain. We use a filter, called the Neural Band-pass Filter (NBP-Filter) to align them to a unified range of frequencies, i.e., spectrum. NBP-Filter learns to pass a specific spectrum while constraining the others. We learn this spectrum in a self-adaptive way to the diffusion time steps, as the noise-level (frequency range) is specific for each step.
% This allows for the mixup of features based on the frequency-affinity and also align them based on the different diffusion steps. 

A bird-eye view of SS-Former is shown in the \ref{fig:framework} (b), which is composed of $N$ blocks that share the same architecture. We set $N=4$ in the paper. Each block consists of two cross-attention-like modules. The first encodes the diffusion noise embedding into the condition semantic embedding, and the next symmetric module encodes the last semantic embedding into the diffusion noise embedding. This allows the model to learn the interaction between noise and semantic features and achieve a stronger representation. Formally, consider $c^{0}$ is the deepest feature embedding of Condition Model and $e$ is that of Diffusion Model. We first transfer $c^{0}$ and $e$ to the Fourier space, denoted as $F(c^{0})$ and $F(e)$, respectively. Note that the feature maps are all patchlized and liner projected in accordance with the standard vision transformer method. Then we compute an affinity weight map over Fourier space taking $e$ as the $query$ and $c^{0}$ as the $key$, which can be represented by $\mathcal{M}=(F(c^{0})\mathcal{W}^{q}) (F(e) \mathcal{W}^{k})^{T}$, where $\mathcal{W}^{q}$ and $\mathcal{W}^{k}$ are the learnable $query$ and $key$ weights in Fourier space. 

We then apply a NBP-Filter to align the representation of frequency. We note that each point in $\mathcal{M}$ now represents a particular frequency, and since we need to control a continuous range of frequencies, it is intuitive to establish a smooth projection from the feature map position to the frequency magnitude. To accomplish this, we use a neural network to learn a weight map from a coordinate map. By doing so, inductive bias of the network will facilitate the learning of a smooth projection, as similar inputs will naturally produce similar outputs\cite{sitzmann2020implicit,wu2019universal}. This idea is widely used in 3D vision tasks and is known as Neural Radiance Fields (NeRF)\cite{Mildenhall20eccv_nerf}. But different from the original NeRF, we further condition it with time-step information. Specifically, the network takes a coordinate map as input and produces an attention map to serve as the filter, both of which have the same size $\mathcal{M}$. We implement it using a simple stack of convolutional blocks with intermediate layer normalization. To condition the network with timestep information, we scale and shift the normalized features with the timestep embedding of the diffusion model. We use two MLP layers to project the current timestep embedding to two values representing the mean and variance, which are used for scaling and shifting, respectively. We stack a total of $R = 6$ such blocks and a Sigmoid function to produce the final filter. Finally, the filter is element-wise multiplied with the affinity map $\mathcal{M}$ in the pipeline. NBP-Filter is trained in an end-to-end manner with the whole pipeline.

The filtered affinity map $\mathcal{M'}$ is then transferred back to Euclidean space using inverse fast Fourier transform (IFFT) and applied to condition features in $value$: $f =  F^{-1}(M') (c^{0} w^{v})$, where $W^{v}$ is the learnable value weights. 
% We then apply the time embedding to an AdaIN normalization following the classic diffusion implementation\cite{nichol2021improved}, which normalizes the feature and then expands it using scale and shift parameters learned from the time embedding. This makes the transformer adaptive to the step information. 
We also use a MLP to further refine the attention result, obtaining the final feature $\tilde{c}^{0}$. The following attention module is symmetric to the first one, but using the combined feature $\tilde{c}^{0}$ as the query and noise embedding $e$ as the key and value, in order to transform the segmentation features to the noise domain. The transformed feature $c^{1}$ will serve as the condition embedding for the next block.

\begin{figure*}
 
    \centering
    \includegraphics[width=\linewidth]{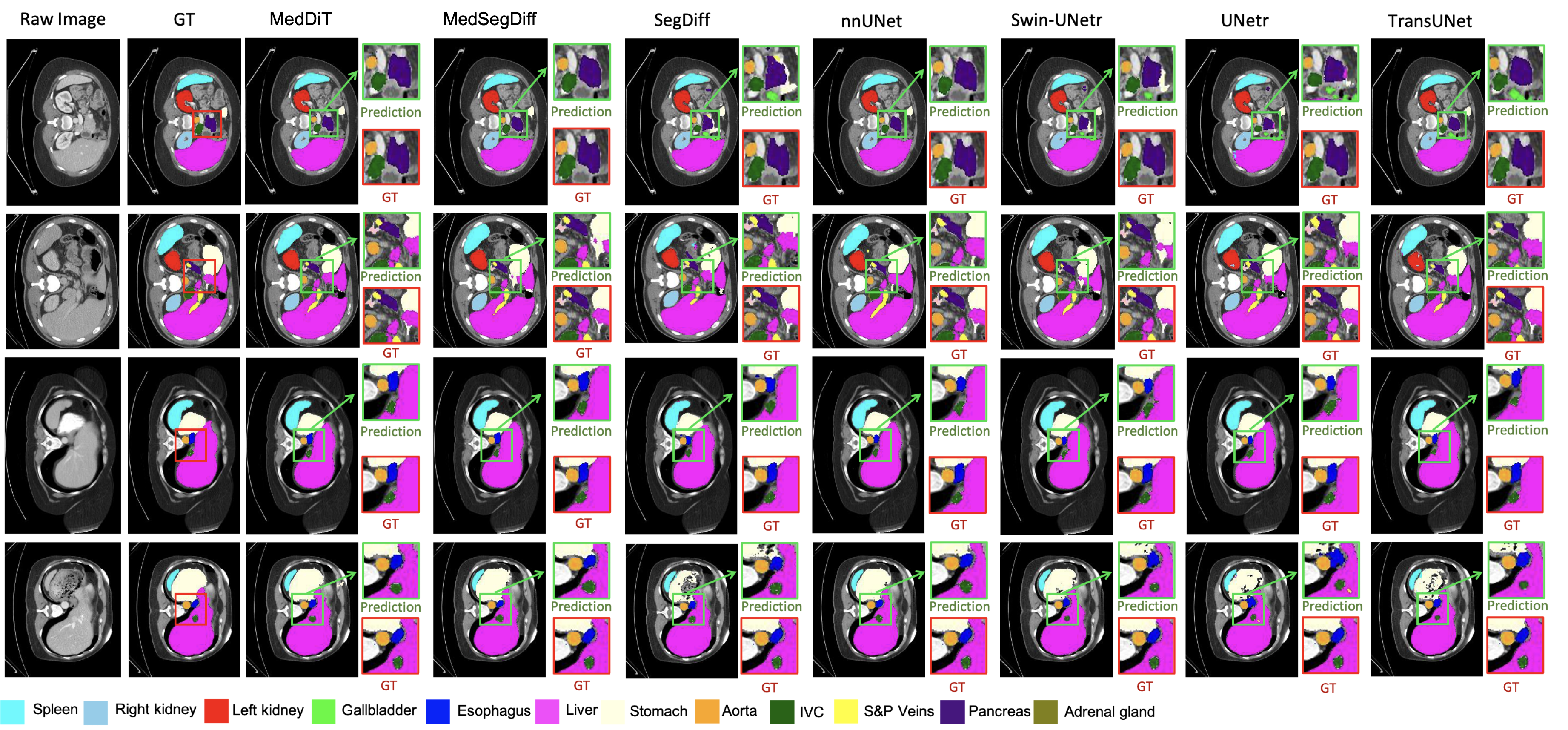}
    \caption{The visual comparison with SOTA segmentation models on BTCV.}
    \label{fig:vis}
     
\end{figure*}

% Figure \ref{fig:vis} presents a qualitative comparison of MedSegDiff-V2 and other competitive methods. It can be observed that MedSegDiff-V2 segments more accurately on parts that are difficult to recognize by the human eye. Due to its ability to benefit from the superior generation capability of the diffusion model and the semantic representation capability of the transformer, it can generate segmentation maps with precise and accurate details, even in low-contrast or ambiguous areas.

% \begin{figure*}
 
%     \centering
%     \includegraphics[width=\linewidth]{multimodal.png}
     
%     \caption{The comparison with SOTA segmentation models on three tasks with different image modalities.}
%     \label{fig:multimodal}
     
% \end{figure*}

\begin{table*}[h]
\centering
 
\caption{The comparison of MedSegDiff-V2 with SOTA segmentation methods on different image modalities. The grey background denotes the methods are proposed for that/these particular tasks.}
\resizebox{0.95\linewidth}{!}{%
\begin{tabular}{c|c|cc|cc|ccc|cc|cc}
\hline
& {\color[HTML]{333333} }   & \multicolumn{2}{c|}{{\color[HTML]{333333} REFUGE2-Disc}}     & \multicolumn{2}{c|}{{\color[HTML]{333333} REFUGE2-Cup}}                                                     & \multicolumn{3}{c|}{{\color[HTML]{333333} BraTs}}                                                  & \multicolumn{2}{c|}{{\color[HTML]{333333} TNMIX}}   & \multicolumn{2}{c}{{\color[HTML]{333333} ISIC}}                                              \\ \hline

& {\color[HTML]{333333} }   & {\color[HTML]{333333} Dice}                         & {\color[HTML]{333333} IoU}     & {\color[HTML]{333333} Dice}                         & {\color[HTML]{333333} IoU}                          & {\color[HTML]{333333} Dice}                         & {\color[HTML]{333333} IoU} & {\color[HTML]{333333} HD95}                          & {\color[HTML]{333333} Dice}                         & {\color[HTML]{333333} IoU} & {\color[HTML]{333333} Dice}                         & {\color[HTML]{333333} IoU} \\ \hline

\multirow{2}{*}{\begin{tabular}[c]{@{}c@{}}Optic\\Disc/Cup\end{tabular}}

& {\color[HTML]{333333} ResUNet}& \cellcolor[HTML]{EFEFEF}{\color[HTML]{333333} 92.9} & \cellcolor[HTML]{EFEFEF}{\color[HTML]{333333} 85.5}  & \cellcolor[HTML]{EFEFEF}{\color[HTML]{333333} 80.1} & \cellcolor[HTML]{EFEFEF}{\color[HTML]{333333} 72.3} & {\color[HTML]{333333} 78.4}                            & {\color[HTML]{333333} 71.3} & {\color[HTML]{333333} 18.71}                             & {\color[HTML]{333333} 78.3}                            & {\color[HTML]{333333} 70.7}      & {\color[HTML]{333333} 87.1}                            & {\color[HTML]{333333} 78.2}                       \\

& {\color[HTML]{333333} BEAL}    & \cellcolor[HTML]{EFEFEF}{\color[HTML]{333333} 93.7} & \cellcolor[HTML]{EFEFEF}{\color[HTML]{333333} 86.1}  & \cellcolor[HTML]{EFEFEF}{\color[HTML]{333333} 83.5} & \cellcolor[HTML]{EFEFEF}{\color[HTML]{333333} 74.1} & {\color[HTML]{333333} 78.8}                            & {\color[HTML]{333333} 71.7} & {\color[HTML]{333333} 18.53}                            & {\color[HTML]{333333} 78.6}                            & {\color[HTML]{333333} 71.6}      & {\color[HTML]{333333} 86.6}                            & {\color[HTML]{333333} 78.0}                       \\ \hline

\multirow{2}{*}{\begin{tabular}[c]{@{}c@{}}Brain\\Tumor\end{tabular}} & {\color[HTML]{333333} TransBTS}  & {\color[HTML]{333333} 94.1}                            & {\color[HTML]{333333} 87.2}   & {\color[HTML]{333333} 85.4}                            & {\color[HTML]{333333} 75.7}                            & \cellcolor[HTML]{EFEFEF}{\color[HTML]{333333} 87.6} & \cellcolor[HTML]{EFEFEF}{\color[HTML]{333333} 78.44}  & \cellcolor[HTML]{EFEFEF}{\color[HTML]{333333} 12.44} & {\color[HTML]{333333} 83.8}                            & {\color[HTML]{333333} 75.5}      & {\color[HTML]{333333} 88.1}                            & {\color[HTML]{333333} 80.6}\\

&  {\color[HTML]{333333} SwinBTS}  & {\color[HTML]{333333} 95.2}                            & {\color[HTML]{333333} 87.7}   & {\color[HTML]{333333} 85.7}                            & {\color[HTML]{333333} 75.9}                            & \cellcolor[HTML]{EFEFEF}{\color[HTML]{333333} 88.7} & \cellcolor[HTML]{EFEFEF}{\color[HTML]{333333} 81.2}  & \cellcolor[HTML]{EFEFEF}{\color[HTML]{333333} 10.03} & {\color[HTML]{333333} 84.5}                            & {\color[HTML]{333333} 76.1}      & {\color[HTML]{333333} 89.8}                            & {\color[HTML]{333333} 82.4}\\
\hline

\multirow{2}{*}{\begin{tabular}[c]{@{}c@{}}Thyroid\\Nodule\end{tabular}} & {\color[HTML]{333333} MTSeg}    & {\color[HTML]{333333} 90.3}                            & {\color[HTML]{333333} 83.6}    & {\color[HTML]{333333} 82.3}                            & {\color[HTML]{333333} 73.1}                            & {\color[HTML]{333333} 82.2}                            & {\color[HTML]{333333} 74.5} & {\color[HTML]{333333} 15.74}                            & \cellcolor[HTML]{EFEFEF}{\color[HTML]{333333} 82.3} & \cellcolor[HTML]{EFEFEF}{\color[HTML]{333333} 75.2} & {\color[HTML]{333333} 87.5} & {\color[HTML]{333333} 79.7} \\

& {\color[HTML]{333333} UltraUNet}  & {\color[HTML]{333333} 91.5}                            & {\color[HTML]{333333} 82.8}  & {\color[HTML]{333333} 83.1}                            & {\color[HTML]{333333} 73.78}                            & {\color[HTML]{333333} 84.5}                            & {\color[HTML]{333333} 76.3}  & {\color[HTML]{333333} 14.03}                            & \cellcolor[HTML]{EFEFEF}{\color[HTML]{333333} 84.5} & \cellcolor[HTML]{EFEFEF}{\color[HTML]{333333} 76.2}  & {\color[HTML]{333333} 89.0}  & {\color[HTML]{333333} 81.8}  \\ \hline

\multirow{2}{*}{\begin{tabular}[c]{@{}c@{}}Skin\\Lesion\end{tabular}} & {\color[HTML]{333333} FAT-Net}    & {\color[HTML]{333333} 91.8}                            & {\color[HTML]{333333} 84.8}    & {\color[HTML]{333333} 80.9}                            & {\color[HTML]{333333} 71.5}                            & {\color[HTML]{333333} 79.2}                            & {\color[HTML]{333333} 72.8} & {\color[HTML]{333333} 17.35}                            & {\color[HTML]{333333} 80.8} & {\color[HTML]{333333} 73.4} & \cellcolor[HTML]{EFEFEF}{\color[HTML]{333333} 90.7} & \cellcolor[HTML]{EFEFEF}{\color[HTML]{333333} 83.9} \\

& {\color[HTML]{333333} BAT}  & {\color[HTML]{333333} 92.3}                            & {\color[HTML]{333333} 85.8}  & {\color[HTML]{333333} 82.0}                            & {\color[HTML]{333333} 73.2}                            & {\color[HTML]{333333} 79.6}                            & {\color[HTML]{333333} 73.5}  & {\color[HTML]{333333} 15.49}                            & {\color[HTML]{333333} 81.7} &{\color[HTML]{333333} 74.2}  & \cellcolor[HTML]{EFEFEF}{\color[HTML]{333333} 91.2}  & \cellcolor[HTML]{EFEFEF}{\color[HTML]{333333} 84.3}  \\ \hline

\multirow{4}{*}{\begin{tabular}[c]{@{}c@{}}General\\ Med Seg\end{tabular}} & nnUNet         & \cellcolor[HTML]{EFEFEF}{\color[HTML]{333333} 94.7}                                                & \cellcolor[HTML]{EFEFEF}{\color[HTML]{333333} 87.3}                 & \cellcolor[HTML]{EFEFEF}{\color[HTML]{333333} 84.9 }                                                & \cellcolor[HTML]{EFEFEF}{\color[HTML]{333333} 75.1}                                                & \cellcolor[HTML]{EFEFEF}{\color[HTML]{333333}88.5}                                                & \cellcolor[HTML]{EFEFEF}{\color[HTML]{333333}80.6}    & \cellcolor[HTML]{EFEFEF}{\color[HTML]{333333}11.20}                                            & \cellcolor[HTML]{EFEFEF}{\color[HTML]{333333}84.2}                                                & \cellcolor[HTML]{EFEFEF}{\color[HTML]{333333}76.2}          & \cellcolor[HTML]{EFEFEF}{\color[HTML]{333333}90.8}                                                & \cellcolor[HTML]{EFEFEF}{\color[HTML]{333333}83.6}                                        \\

& TransUNet          & \cellcolor[HTML]{EFEFEF}{\color[HTML]{333333}95.0}                                                & \cellcolor[HTML]{EFEFEF}{\color[HTML]{333333}87.7}                        & \cellcolor[HTML]{EFEFEF}{\color[HTML]{333333}85.6}                                                & \cellcolor[HTML]{EFEFEF}{\color[HTML]{333333}75.9}                                                & \cellcolor[HTML]{EFEFEF}{\color[HTML]{333333}86.6}                                                & \cellcolor[HTML]{EFEFEF}{\color[HTML]{333333}79.0}       & \cellcolor[HTML]{EFEFEF}{\color[HTML]{333333}13.74}                                          & \cellcolor[HTML]{EFEFEF}{\color[HTML]{333333} 83.5}                         & \cellcolor[HTML]{EFEFEF}{\color[HTML]{333333}75.1}        & \cellcolor[HTML]{EFEFEF}{\color[HTML]{333333}89.4}       & \cellcolor[HTML]{EFEFEF}{\color[HTML]{333333}82.2}                                            \\

& {\color[HTML]{333333} UNetr}   & \cellcolor[HTML]{EFEFEF}{\color[HTML]{333333}94.9}                                                &  \cellcolor[HTML]{EFEFEF}{\color[HTML]{333333}87.5}  & \cellcolor[HTML]{EFEFEF}{\color[HTML]{333333}83.2}                         & \cellcolor[HTML]{EFEFEF}{\color[HTML]{333333} 73.3}                         & \cellcolor[HTML]{EFEFEF}{\color[HTML]{333333} 87.3}                         & \cellcolor[HTML]{EFEFEF}{\color[HTML]{333333} 80.6}  & \cellcolor[HTML]{EFEFEF}{\color[HTML]{333333}12.81}                       & \cellcolor[HTML]{EFEFEF}{\color[HTML]{333333} 81.7}                         & \cellcolor[HTML]{EFEFEF}{\color[HTML]{333333} 73.5}          & \cellcolor[HTML]{EFEFEF}{\color[HTML]{333333}89.7}                                                & \cellcolor[HTML]{EFEFEF}{\color[HTML]{333333}82.8}                 \\

& {\color[HTML]{333333} Swin-UNetr}    & \cellcolor[HTML]{EFEFEF}{\color[HTML]{333333}95.3}                                                & \cellcolor[HTML]{EFEFEF}{\color[HTML]{333333}87.9}   &\cellcolor[HTML]{EFEFEF}{\color[HTML]{333333}84.3}                         & \cellcolor[HTML]{EFEFEF}{\color[HTML]{333333}74.5}                         & \cellcolor[HTML]{EFEFEF}{\color[HTML]{333333} 88.4}                         & \cellcolor[HTML]{EFEFEF}{\color[HTML]{333333} 81.8}               & \cellcolor[HTML]{EFEFEF}{\color[HTML]{333333}11.36}           & \cellcolor[HTML]{EFEFEF}{\color[HTML]{333333}83.5}                         & \cellcolor[HTML]{EFEFEF}{\color[HTML]{333333} 74.8}        & \cellcolor[HTML]{EFEFEF}{\color[HTML]{333333}90.2}                                                & \cellcolor[HTML]{EFEFEF}{\color[HTML]{333333}83.1}                   \\ \hline

\multirow{5}{*}{\begin{tabular}[c]{@{}c@{}} Diffusion\\ Based\end{tabular}} & {\color[HTML]{333333} EnsemDiff}  & \cellcolor[HTML]{EFEFEF}{\color[HTML]{333333} 94.3}                            & \cellcolor[HTML]{EFEFEF}{\color[HTML]{333333} 87.8}  & \cellcolor[HTML]{EFEFEF}{\color[HTML]{333333} 84.2}                            & \cellcolor[HTML]{EFEFEF}{\color[HTML]{333333} 74.4}                            & \cellcolor[HTML]{EFEFEF}{\color[HTML]{333333} 88.7} & \cellcolor[HTML]{EFEFEF}{\color[HTML]{333333} 80.9} & \cellcolor[HTML]{EFEFEF}{\color[HTML]{333333} 10.85} & \cellcolor[HTML]{EFEFEF}{\color[HTML]{333333} 83.9}                            & \cellcolor[HTML]{EFEFEF}{\color[HTML]{333333} 75.3}       & \cellcolor[HTML]{EFEFEF}{\color[HTML]{333333} 88.2}                            & \cellcolor[HTML]{EFEFEF}{\color[HTML]{333333} 80.7}                       \\ 

& {\color[HTML]{333333} SegDiff}   & \cellcolor[HTML]{EFEFEF}{\color[HTML]{333333} 92.6}                         & \cellcolor[HTML]{EFEFEF}{\color[HTML]{333333} 85.2}         & \cellcolor[HTML]{EFEFEF}{\color[HTML]{333333}82.5}                         & \cellcolor[HTML]{EFEFEF}{\color[HTML]{333333} 71.9}                         & \cellcolor[HTML]{EFEFEF}{\color[HTML]{333333} 85.7}                         & \cellcolor[HTML]{EFEFEF}{\color[HTML]{333333} 77.0}     & \cellcolor[HTML]{EFEFEF}{\color[HTML]{333333}14.31 }                  & \cellcolor[HTML]{EFEFEF}{\color[HTML]{333333} 81.9}                         & \cellcolor[HTML]{EFEFEF}{\color[HTML]{333333}74.8}   & \cellcolor[HTML]{EFEFEF}{\color[HTML]{333333} 87.3}                         & \cellcolor[HTML]{EFEFEF}{\color[HTML]{333333} 79.4}                         \\
 
& {\color[HTML]{333333} MedsegDiff}  & \cellcolor[HTML]{EFEFEF}{\color[HTML]{333333} 95.1}                         & \cellcolor[HTML]{EFEFEF}{\color[HTML]{333333} 87.6}     & \cellcolor[HTML]{EFEFEF}{\color[HTML]{333333} 85.9}                         & \cellcolor[HTML]{EFEFEF}{\color[HTML]{333333} 76.2}                         & \cellcolor[HTML]{EFEFEF}{\color[HTML]{333333} 88.9}                         & \cellcolor[HTML]{EFEFEF}{\color[HTML]{333333} 81.2}     & \cellcolor[HTML]{EFEFEF}{\color[HTML]{333333} 10.41}                    & \cellcolor[HTML]{EFEFEF}{\color[HTML]{333333} 84.8}                         & \cellcolor[HTML]{EFEFEF}{\color[HTML]{333333} 76.4}           & \cellcolor[HTML]{EFEFEF}{\color[HTML]{333333} 91.3}                         & \cellcolor[HTML]{EFEFEF}{\color[HTML]{333333} 84.1}                 \\

% & {\color[HTML]{333333} AmbDiff}   & \cellcolor[HTML]{EFEFEF}{\color[HTML]{333333} 92.8}                         & \cellcolor[HTML]{EFEFEF}{\color[HTML]{333333} 85.5}         & \cellcolor[HTML]{EFEFEF}{\color[HTML]{333333}82.9}                         & \cellcolor[HTML]{EFEFEF}{\color[HTML]{333333} 72.6}                         & \cellcolor[HTML]{EFEFEF}{\color[HTML]{333333} 86.3}                         & \cellcolor[HTML]{EFEFEF}{\color[HTML]{333333} 77.8}     & \cellcolor[HTML]{EFEFEF}{\color[HTML]{333333}12.86 }                  & \cellcolor[HTML]{EFEFEF}{\color[HTML]{333333} 82.8}                         & \cellcolor[HTML]{EFEFEF}{\color[HTML]{333333}75.9}   & \cellcolor[HTML]{EFEFEF}{\color[HTML]{333333} 89.5}                         & \cellcolor[HTML]{EFEFEF}{\color[HTML]{333333} 79.7}                         \\

& {\color[HTML]{333333} MedsegDiff+TransUNet} & \cellcolor[HTML]{EFEFEF}{\color[HTML]{333333} 91.8}                         & \cellcolor[HTML]{EFEFEF}{\color[HTML]{333333} 84.5}   & \cellcolor[HTML]{EFEFEF}{\color[HTML]{333333} 82.1}                         & \cellcolor[HTML]{EFEFEF}{\color[HTML]{333333} 72.6}                         & \cellcolor[HTML]{EFEFEF}{\color[HTML]{333333} 86.1}                         & \cellcolor[HTML]{EFEFEF}{\color[HTML]{333333} 78.0}            & \cellcolor[HTML]{EFEFEF}{\color[HTML]{333333}13.88}             & \cellcolor[HTML]{EFEFEF}{\color[HTML]{333333} 79.2}                         & \cellcolor[HTML]{EFEFEF}{\color[HTML]{333333} 71.4}     & \cellcolor[HTML]{EFEFEF}{\color[HTML]{333333} 84.6}                         & \cellcolor[HTML]{EFEFEF}{\color[HTML]{333333} 75.5}                     \\
\hline

Proposed & MedSegDiff-V2             & \cellcolor[HTML]{EFEFEF}{\color[HTML]{333333}\textbf{96.7}}                                       & \cellcolor[HTML]{EFEFEF}{\color[HTML]{333333}\textbf{88.9}}           & \cellcolor[HTML]{EFEFEF}{\color[HTML]{333333}\textbf{87.9}}                                       & \cellcolor[HTML]{EFEFEF}{\color[HTML]{333333}\textbf{80.3}}                                       & \cellcolor[HTML]{EFEFEF}{\color[HTML]{333333}\textbf{90.8}}                                       & \cellcolor[HTML]{EFEFEF}{\color[HTML]{333333}\textbf{83.4}}       & \cellcolor[HTML]{EFEFEF}{\color[HTML]{333333}\textbf{7.53}}                                 & \cellcolor[HTML]{EFEFEF}{\color[HTML]{333333}\textbf{88.7}}                                       & \cellcolor[HTML]{EFEFEF}{\color[HTML]{333333}\textbf{81.5}}      & \cellcolor[HTML]{EFEFEF}{\color[HTML]{333333}\textbf{93.2}}                                       & \cellcolor[HTML]{EFEFEF}{\color[HTML]{333333}\textbf{85.3}}                                  \\ \hline
\end{tabular}}\label{tab:muti}
 
\end{table*}

\begin{table*}[h]
\centering
 
\caption{The comparison of MedSegDiff-V2 with SOTA segmentation methods over AMOS dataset evaluated by Dice Score. Best results are denoted as \textbf{bold}.}
\resizebox{\linewidth}{!}{%
\begin{tabular}{c|ccccccccccccccc|c}
\hline
Methods                                                      & Spleen         & R.Kid          & L.Kid          & Gall.          & Eso.           & Liver          & Stom.          & Aorta          & IVC            & Panc.          & RAG            & LAG            & Duo.           & Blad.          & Pros.          & Avg            \\ \hline
TransUNet                                                    & 0.881          & 0.928          & 0.919          & 0.813          & 0.740          & 0.973          & 0.832          & 0.919          & 0.841          & 0.713          & 0.638          & 0.565          & 0.685          & 0.748          & 0.692          & 0.792          \\ 

UNetr                                                        & 0.926          & 0.936          & 0.918          & 0.785          & 0.702          & 0.969          & 0.788          & 0.893          & 0.828          & 0.732          & 0.717          & 0.554          & 0.658          & 0.683          & 0.722          & 0.762          \\ 
Swin-UNetr                                                   & 0.959          & 0.960          & 0.949          & 0.894          & 0.827          & \textbf{0.979} & 0.899          & 0.944          & 0.899          & 0.828          & 0.791          & 0.745          & 0.817          & \textbf{0.875} & 0.841          & 0.880          \\ 
nnUNet                                                       & 0.965         & 0.959          & 0.951          & 0.889          & 0.820          & 0.980          & 0.890          & 0.948          & 0.901          & 0.821          & 0.785          & 0.739          & 0.806          & 0.869          & 0.839          & 0.878          \\ 
\begin{tabular}[c]{@{}c@{}}EnsDiff\end{tabular} & 0.905          & 0.918          & 0.904          & 0.732          & 0.723          & 0.947          & 0.838          & 0.915          & 0.838          & 0.704          & 0.677          & 0.618          & 0.715          & 0.673          & 0.680          & 0.786          \\ 
SegDiff & 0.885          & 0.872          & 0.891          & 0.703          & 0.654          & 0.852          & 0.702          & 0.874          & 0.819          & 0.715          & 0.654          & 0.632          & 0.697          & 0.652          & 0.695          & 0.753          \\ 
MedSegDiff                                                   & 0.963          & 0.965          & 0.953          & 0.917          & 0.846          & 0.971          & 0.906          & 0.952          & 0.918          & 0.854          & 0.803          & 0.751          & 0.819          & 0.868          & 0.855          & 0.889          \\ 
\begin{tabular}[c]{@{}c@{}}MedSegDiff \\ + TransUNet\end{tabular} & 0.941          & 0.932          & 0.921          & 0.934          & 0.813          & 0.946          & 0.867          & 0.921          & 0.880          & 0.821          & 0.793          & 0.528          & 0.788          & 0.813          & 0.837          & 0.849          \\ \hline
Anchor                                                   & 0.872          & 0.901          & 0.892          & 0.784          & 0.802          & 0.910          & 0.835          & 0.908          & 0.810          & 0.735          & 0.682          & 0.651          & 0.583          & 0.631          & 0.728          & 0.781          \\ 
MedSegDiff-V2                                                & \textbf{0.971} & \textbf{0.969} & \textbf{0.964} & \textbf{0.932} & \textbf{0.864} & 0.976          & \textbf{0.934} & \textbf{0.968} & \textbf{0.925} & \textbf{0.871} & \textbf{0.815} & \textbf{0.762} & \textbf{0.827} & 0.873          & \textbf{0.871} & \textbf{0.901} \\ \hline
\end{tabular}%
}\label{tab:amos}
 
\end{table*}

\begin{table*}[h]
\centering
 
\caption{The comparison of MedSegDiff-V2 with SOTA segmentation methods over BTCV dataset evaluated by Dice Score. Best results are denoted as \textbf{bold}.}
\resizebox{0.9\linewidth}{!}{%
\begin{tabular}{c|cccccccccccc|c}
\hline
Model                                                           & Spleen & R.Kid & L.Kid & Gall. & Eso.  & Liver & Stom.  & Aorta & IVC  &Veins & Panc. & AG  & Ave  \\ \hline
TransUNet                                                       & 0.952                        & 0.927 & 0.929 & 0.662 & 0.757 & 0.969  & 0.889 & 0.920  & 0.833 & 0.791       & 0.775     & 0.637 & 0.838 \\
UNetr                                                           & 0.968                        & 0.924 & 0.941 & 0.750 & 0.766 & 0.971  & 0.913 & 0.890  & 0.847 & 0.788       & 0.767     & 0.741 & 0.856 \\
Swin-UNetr                                                      & 0.971                        & 0.936 & 0.943 & 0.794 & 0.773 & 0.975  & 0.921 & 0.892  & 0.853 & 0.812       & 0.794     & 0.765 &  0.869    \\
nnUNet                                                          & 0.942                        & 0.894 & 0.910 & 0.704 & 0.723 & 0.948  & 0.824 & 0.877  & 0.782 & 0.720       & 0.680     & 0.616 & 0.802 \\
EnsDiff                                                         & 0.938                        & 0.931 & 0.924 & 0.772 & 0.771 & 0.967  & 0.910 & 0.869  & 0.851 & 0.802       & 0.771     & 0.745 & 0.854      \\
SegDiff                                                         & 0.954                        & 0.932 & 0.926 & 0.738 & 0.763 & 0.953  & 0.927 & 0.846  & 0.833 & 0.796       & 0.782     & 0.723 & 0.847     \\
MedSegDiff                                                      & 0.973                        & 0.930 & 0.955 & 0.812 & 0.815 & 0.973  & 0.924 & 0.907  & 0.868 & 0.825       & 0.788     & 0.779 & 0.879     \\
\begin{tabular}[c]{@{}c@{}}MedSegDiff\\ +TransUNet\end{tabular} & 0.912                        & 0.876 & 0.846 & 0.645 & 0.718 & 0.947  & 0.824 & 0.876  & 0.715 & 0.775       & 0.672     & 0.618 & 0.785     \\ \hline
Anchor                                                          & 0.928                        & 0.882 & 0.873 & 0.652 & 0.750 & 0.951  & 0.829 & 0.855  & 0.731 & 0.714       & 0.683     & 0.602 & 0.787     \\
MedSegDiff-V2                                                   & \textbf{0.978}                        & \textbf{0.941} & \textbf{0.963} & \textbf{0.848} & \textbf{0.818} & \textbf{0.985}  & \textbf{0.940} & \textbf{0.928}  & \textbf{0.869} & \textbf{0.823}       & \textbf{0.831}     & \textbf{0.817} &  \textbf{0.895}   \\ \hline
\end{tabular}}\label{tab:btcv}
 
\end{table*}

\begin{table*}[h]
\centering
 
\caption{An ablation study on Anchor Conditioning and SS-Former. SA denotes Spatial Attention.}
\resizebox{0.9\linewidth}{!}{%
\begin{tabular}{c|c|c|c|ccccc}
\hline
\multicolumn{2}{c|}{ Anc.Cond.} & \multicolumn{2}{c|}{Sem.Cond.} & AMOS & BTCV & OpticCup & BrainTumor & ThyroidNodule \\ \hline
SA & $\mathcal{U}$-SA & \begin{tabular}{@{}c@{}}SS-Former \\  (w/o Filter)\end{tabular} & NBP-Filter & Ave-Dice (\%) & Ave-Dice (\%)  & Dice (\%)& Dice (\%)&Dice (\%) \\ \hline
 & & &              & 78.6 & 85.4     & 84.6     & 88.2       & 84.1          \\ 
\checkmark &  & &  & 83.5 & 85.8  & 85.2     & 88.7       & 84.6 \\ 
&\checkmark    &     &      &  86.7  & 86.6   & 85.7     & 89.4       & 86.5          \\
&\checkmark   &  \checkmark  &       & 87.8 & 87.1     & 86.5     & 89.8       & 86.8          \\
&\checkmark   &  \checkmark  &  \checkmark      & \textbf{90.1} & \textbf{89.5}   & \textbf{87.9} & \textbf{90.8} & \textbf{88.7} \\ \hline     
\end{tabular}}\label{tab:ab}
 
\end{table*}

\section{Experiments}\label{sec:exp}
  
\subsection{Dataset}
We conduct the experiments on total five different medical image segmentation datasets. Two datasets are used to verify the general segmentation performance, which are public AMOS2022\cite{ji2022amos} dataset with sixteen anatomies and public BTCV\cite{fang2020multi} dataset with twelve anatomies annotated for abdominal multi-organ segmentation. The other four public datasets REFUGE-2 \cite{fang2022REFUGE2}, BraTs-2021 dataset \cite{baid2021rsna}, ISIC 2018 dataset\cite{milton2019automated} and TNMIX dataset \cite{pedraza2015open} are used to verify the model performance on multi-modal images, which are the optic-cup segmentation from fundus images, the brain tumor segmentation from MRI images, and the thyroid nodule segmentation from ultrasound images. More details about datasets are shown in the appendix.

%The experiments of glaucoma, thyroid cancer and melanoma diagnosis are conducted on , which contain 1200, 2000, 8046 samples, respectively. The datasets are publicly available with segmentation labels. Train/validation/test sets are split following the default settings of the dataset.

\subsection{Implementation Details}
  
All experiments were conducted using the PyTorch platform and trained/tested on 4 NVIDIA A100 GPUs. All images were uniformly resized to a resolution of 256$\times$256 pixels. The networks were trained in an end-to-end manner using the AdamW\cite{loshchilov2017decoupled} optimizer with a batch size of 32. The initial learning rate was set to 1 $\times 10^{-4}$. We employed 100 diffusion steps for the inference. We run the model 10 times for the ensemble, which is much fewer than the 25 times in MedSegDiff\cite{wu2022medsegdiff}. Then we use STAPLE algorithm\cite{warfield2004simultaneous} to fuse the different samples. We evaluate the segmentation performance by Dice score, IoU, HD95 metrics.

\subsection{Main Results}
\subsubsection{Comparing with SOTA on Abdominal Multi-organ Segmentation}
  
To verify the general medical image segmentation performance, we compare MedSegDiff-V2 with SOTA segmentation methods on multi-organ segmentation dataset AMOS and BTCV. The quantitative results of Dice score are shown in \ref{tab:amos} and \ref{tab:btcv} respectively. In the table, we compare with the segmentation methods which are widely-used and well-recognized in the community, including the CNN-based method nnUNet\cite{isensee2021nnu}, the transformer-based methods TransUNet\cite{chen2021transunet}, UNetr\cite{hatamizadeh2022unetr}, Swin-UNetr\cite{jiang2022swinbts} and the diffusion based method EnsDiff \cite{wolleb2021diffusion}, SegDiff\cite{amit2021segdiff},  MedSegDiff \cite{wu2022medsegdiff}. We also compare with a simple combination of diffusion and transformer model. We replace the UNet model in MedSegDiff to TransUNet and denoted it as 'MedSegDiff + TransUNet' in the table. 
%The detailed settings of the compared methods are shown in the appendix.

As seen in \ref{tab:amos} and \ref{tab:btcv}, advanced network architectures and sophisticated designs are crucial for achieving good performance. Considering the architecture, transformer-based models such as Swin-UNetr outperform the carefully designed CNN-based model, nnUNet. The diffusion-based model MedSegDiff again outperforms the transformer-based models on most of the organs. However, network architecture alone is not the determining factor for performance. For example, the well-designed CNN-based model nnUNet considerably outperforms the transformer-based model TransUNet and UNetr in the table. This is also true for diffusion-based models. We can see that a straightforward adoption of the diffusion model for medical image segmentation, i.e., EnsDiff and SegDiff, perform worse than UNetr and Swin-UNetr. A simple combination of transformer and diffusion model, i.e., MedSegDiff + TransUNet, obtains even worse performance than the standard MedSegDiff. 
%This is because the transformer is more sensitive to adaptive conditions and extracts more delicate semantic features that diverge from the diffusion backbone. 
By introducing Anchor Condition and SS-Former in the diffusion + transformer model, MedSegDiff-V2 overcomes these challenges and shows superior performance. We also present a qualitative comparison in \ref{fig:vis}. It can be observed that MedSegDiff-V2 predicts segmentation maps with more precise details, even in low-contrast or ambiguous areas.
% We compare it with diffusion-based models, i.e., EnsDiff and MedSegDiff, using the same ensemble times (all set to five times), and it produces more stable and accurate results as shown in the table.

\subsubsection{Comparing with SOTA on Multi-modality Images}
We also compare MedSegDiff-V2 to SOTA segmentation methods proposed for three specific tasks with different image modalities. The results are presented in \ref{tab:muti}. In the table, ResUnet\cite{yu2019robust} and BEAL\cite{wang2019boundary} are proposed for optic cup segmentation, TransBTS\cite{wang2021transbts} and SwinBTS\cite{wang2021transbts} are proposed for brain tumor segmentation, MTSeg\cite{gong2021multi} and UltraUNet\cite{chu2021ultrasonic} are proposed for thyroid nodule segmentation, and FAT-Net\cite{wu2022fat} and BAT\cite{wang2021boundary} are proposed for skin lesion segmentation. 
%We also compare to general medical image segmentation methods on these three datasets. 

From the table, we can see that MedSegDiff-V2 surpasses all other methods in five different tasks, highlighting its remarkable generalization capability across various medical segmentation tasks and image modalities. In comparison to the UNet-based MedSegDiff, MedSegDiff-V2 exhibits improvements of 2.0\% on Optic-Cup, 1.9\% on Brain-Tumor, and 3.9\% on Thyroid Nodule in terms of the Dice score, underscoring the effectiveness of its transformer-based backbone. Furthermore, when compared to MedSegDiff plus TransUNet, MedSegDiff-V2 outperforms it by an even larger margin, clearly demonstrating the efficacy of the proposed $\mathcal{U}$-SA  and SS-Former in enhancing performance.

\subsubsection{Ablation Study}
We conducted a comprehensive ablation study to verify the effectiveness of the proposed modules. The results are shown in \ref{tab:ab}, where Anc.Cond. and Sem.Cond. denote Anchor Condition and Semantic Condition, respectively. As shown in the table, Anc.Cond. significantly improves the vanilla diffusion model, with the proposed $\mathcal{U}$-SA outperforming the previous Spatial Attention on all datasets. In Sem.Cond., using SS-Former alone provides only marginal improvement, but combining it with the NBP-Filter results in a significant improvement, demonstrating the effectiveness of the proposed SS-Former design.

  \begin{figure}[h]
    \centering
    \includegraphics[width=\linewidth]{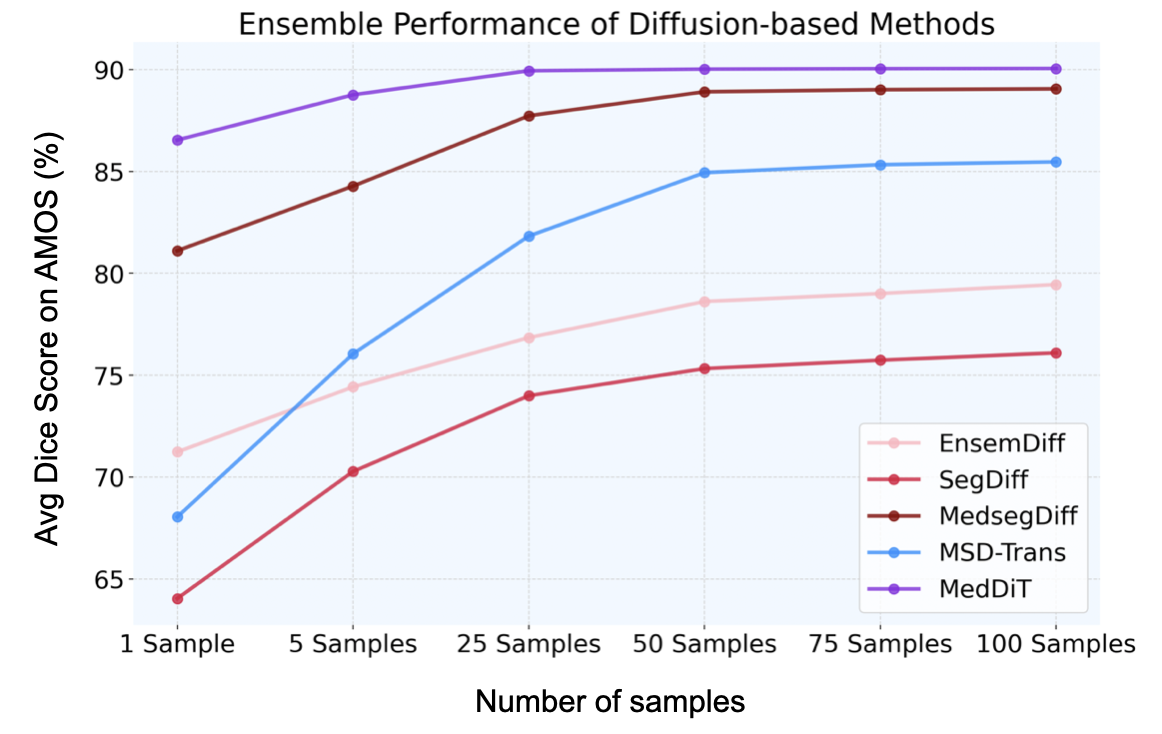}
    \caption{The comparison of ensemble effect of DPM-based methods. We show their performance of average Dice Score on AMOS with increasing sampling times.}
    \label{fig:converge}
\end{figure}

\subsection{Analysis and Discussion}

\subsubsection{Implicit Ensemble Effect}
As confirmed by numerous previous studies \cite{wolleb2021diffusion, wu2022medsegdiff, amit2021segdiff}, the implicit ensemble of multiple sampling runs plays a crucial role in diffusion-based methods. In diffusion model context, implicit ensemble refers to combining predictions from multiple samplings of a single diffusion model, rather than fusing predictions from different models.

In this study, we evaluate the ensemble performance of various diffusion-based medical segmentation models, as shown in \ref{fig:converge}. The evaluation is based on the average Dice Score calculated on the AMOS dataset. Each configuration is run 20 times, and the average Dice Score is used as the performance metric. In the figure, we denote "MedSegDiff+TransUNet" setting as "MSD-Trans". Our findings indicate a common trend, where the model performance improves rapidly in the initial 50 ensembles and then stabilizes. Typically, the best performance is achieved after approximately 50 ensembles.

When comparing MedSegDiff-V2 variant with other diffusion methods, we observe that it requires fewer ensembles to converge. It starts with a significantly better performance, surpassing MedSegDiff by 5\%, and consistently maintains a lead of over 2\% throughout. This highlights the efficiency of MedSegDiff-V2, as it achieves satisfactory results even with fewer ensemble iterations. Moreover, it suggests that a superior starting point and more stable predictions can lead to a higher performance ceiling. This aligns with our assumption that low-quality samples can consistently degrade the model's performance by introducing noise. This again demonstrates the importance of introducing $\mathcal{U}$-SA for divergence control and utilizing SS-Former to attain a better starting point.

\subsubsection{Analysis of Uncertainty}
In \ref{tab:analysis}, we compare the sample diversity on REFUGE2-Cup dataset. We compare the previous DPM-based methods, backbone with individual proposed modules, and final MedSegDiff-V2 together. We evaluated the variance among the samples using the Generalized Energy Distance (GED) and Confidence Interval (CI). GED is a commonly used metric to measures the agreement between predictions and the ground truth distribution of segmentation by comparing their distributions\cite{kohl2018probabilistic}. A lower energy value indicates better agreement.

From the table, we can see that the proposed $\mathcal{U}$-SA achieves lower CI and higher GED compared to previous methods, indicating a larger sample diversity. However, it is also observed that the proposed model reaches a higher or comparable performance, suggesting that its generated samples mostly fall within the uncertainty region of the targets. When using SS-Former alone, without $\mathcal{U}$-SA, the model achieves the best agreement with the highest CI and lowest GED. Although SS-Former gets a fine performance with largest confidence, it fails to fully use the diversity ensemble capability of the diffusion model. By combining $\mathcal{U}$-SA and SS-Former as MedSegDiff-V2, the performance is significantly improved with still high confidence. It suggests that SS-Former helps mitigate the noise generated in $\mathcal{U}$-SA, while $\mathcal{U}$-SA provides more diversity to the model, resulting in mutual improvement.

\subsubsection{Model Efficiency and Complexity}
In \ref{tab:analysis}, we also present a comparison of model complexity and Gflops with other diffusion-based segmentation methods. The reported Gflops is the processing speed for a single $256 \times 256$ image until stability is reached in the implicit ensemble. We consider a variance of performance less than 0.1\% across the last ten ensembles as an indicator of convergence. This metric is significant for the practical application of diffusion-based segmentation models, as users commonly run the diffusion model iteratively to obtain a stable result.

We can see from the table that, unlike traditional deep learning models, the amount of parameters in diffusion-based models is not directly correlated with Gflops, due to the presence of the implicit ensemble. For instance, even though MedSegDiff-V2 incorporates transformer blocks and occupies more parameters, it requires fewer Gflops as it achieves stability in fewer steps. In comparison, MedSegDiff-V2 consumes only half the Gflops of MedSegDiff while outperforming it in various segmentation tasks, as demonstrated above. This underscores the efficiency of MedSegDiff-V2 and its potential in real-world application.

% Despite having fewer parameters than traditional transformer-based models, MedSegDiff-V2, like the other diffusion-based models, exhibits significantly longer inference time. This is a inherent limitation of all diffusion-based methods and remains a significant challenge within the research community.

% To address this issue, we explored several recently proposed DPM accelerating algorithms \cite{lu2022dpm, liu2022flow, watson2022learning} for MedSegDiff-V2. However, we discovered that these algorithms did not perform as effectively in our conditional-DPM segmentation setting as they did in image generation cases. This suggests that Conditional-DPM variants may require specific acceleration algorithms tailored to their unique characteristics. This intriguing finding motivates us to delve deeper into this topic in our future work, where we plan to investigate and explore potential solutions.

\begin{table}[h]
\centering
% \caption{Comparison of the generated sample uncertainty. Diffusion backbone with exclusively $\mathcal{U}$-SA and SS-Former are also compared to check their individual effects. }
% \begin{minipage}{.47\linewidth}
\caption{Comparison of model parameters, Gflops, and generated samples uncertainty}
\resizebox{\columnwidth}{!}{%
\begin{tabular}{c|cc|ccc}
\hline
     Model     & Params (M) & Gflops    & CI & GED & Dice\\ \hline
% ResUNet      & 12         & 0.02                     \\
% MRNet        & 48         & 0.09                     \\
% nnUNet       & 19         & 0.03                     \\
% TransUNet    & 96         & 0.08                     \\ 
% UNetr    & 104         & 0.12                     \\ 
% Swin-UNetr    & 138         & 0.15                     \\ \hline
EnsemDiff & \textbf{23}         & 2203 & 76.3    & 28.9      &  84.2            \\
SegDiff  & \textbf{23}         & 2399 & 75.4    & 26.4   &  82.5  \\   
MedSegDiff  & 25         & 1770 & 77.5    & 27.9  &  85.9 \\ 
MSD-Trans & 118         & 2581 & 75.8    & 28.7  &  82.1 \\ \hline 
bone + $\mathcal{U}$-SA  & -         & - & 73.2    & 34.6  &  85.7 \\
bone + SS-Former & -         & - & \textbf{84.2}    & \textbf{21.7}  &  86.1 \\
MedSegDiff-V2 & 46       & \textbf{983} & 82.6    & 23.5  &  \textbf{87.9}    \\ \hline 
\end{tabular}}\label{tab:analysis}
\end{table}
  
\section{Conclusion}
  
In this paper, we enhance the diffusion-based medical image segmentation framework by incorporating the transformer mechanism into the original UNet backbone, called MedSegDiff-V2. We propose a novel SS-Former architecture to learn the interaction between noise and semantic features. The comparative experiments show our model outperformed previous SOTA methods on 20 different medical image segmentation tasks with various image modalities. As the first transformer-based diffusion model for medical image segmentation, we believe MedSegDiff-V2 will serve as a benchmark for future research.
\clearpage

\bibliography{aaai24}

\end{document}